\documentclass[pra,nofootinbib,aps,reprent,showpacs,amsmath,longbibliography]{revtex4}
\usepackage{graphicx}
\usepackage{epstopdf} 
\usepackage{color}
\usepackage{amsmath, amssymb}

\newcounter{mycount}

\newcommand{\be}[1]{ \begin{eqnarray} \mbox{$\label{#1}$} }
   
\newcommand{\ee}{\end{eqnarray}}
\newcommand{\eeq}{\end{equation}}

\newcommand{\pref}[1]{(\ref{#1})}

\newcommand\ie {{\it i.e. }}

\newcommand\etc{{\it etc. }}

\newcommand\ket [1] {|#1 \rangle }

\newcommand{\av}[1]{\langle #1\rangle}

\newcommand{\oncitec}[1]{Ref.\,\onlinecite{#1}, }

\begin{document}

\title{Screening and Topological Order in Thin Superconducting Films }

\author{E. C. Marino}

\author{D. Niemeyer}

\affiliation{Instituto de F\'\i sica, Universidade Federal do Rio de Janeiro, Rio de Janeiro RJ, 21941-972, Brazil}

\author{Van S\'ergio Alves}

\affiliation{Faculdade de F\'\i sica, Universidade Federal do Par\'a, Bel\'em, 66075-110 PA, Brazil}

\author{T. H. Hansson }

\affiliation{Department of Physics, Stockholm University, AlbaNova University Center, SE-106 91 Stockholm, Sweden }
\affiliation{Nordita,
KTH Royal Institute of Technology and Stockholm University,
Roslagstullsbacken 23, SE-106 91 Stockholm, Sweden}

\author{Sergej Moroz}
\affiliation{Department of Physics, Technical University of Munich, 85748 Garching, Germany}

\date{\today}

\begin{abstract} 

We derive an effective two-dimensional low-energy theory for thin superconducting films coupled to a three-dimensional fluctuating electromagnetic field. Using this theory we discuss plasma oscillations, interactions between charges and vortices and extract the energy of a vortex. Having found that the effective theory properly describes the long distance physics, we then use it to investigate to what extent the superconducting film is a topologically ordered phase of matter.

\end{abstract}

\maketitle


\section{Introduction}
Since the work of Wen \cite{wen1990topological} it is known that an ordinary fluctuating superconductor is an example of a topologically ordered phase with characteristic features such as unusual charges, nontrivial braiding statistics and  topological ground state degeneracy on a torus. As behooves a topological phase, the leading term in the low-energy effective action is a topological field theory\cite{bala92}, and more specifically a BF theory \cite{blau1991topological}.\footnote{A recent study \cite{Moroz2017} indicates that the complete topological theory of an s-wave superconductor is not a purely bosonic BF theory, but also contains elementary fermions that braid trivially with all excitations.} Although the BF term is present in any dimension, the nature of the excitations, which are vortices and quasiparticles, differ. In two spatial dimensions, which will be the subject of this paper, the vortices are pointlike, and the nontrivial statistical braiding phase is simply the minus sign that the wave function acquires as a Bogoliubov quasiparticle  encircles   a vortex.

The effective action for an idealized two-dimensional superconductor, which was derived in \cite{Hansson:2004fk}, contains, in addition to the BF term ${\mathcal L}_{BF} = (1/\pi)\epsilon^{\mu\nu\sigma}a_\mu\partial_\nu b_\sigma$, also Maxwell terms for the  two-dimensional gauge fields $a_\mu$ and $b_\mu$. 
The starting point there was a relativistic two-dimensional abelian Higgs model coupled to QED$_3$, \ie two-dimensional (2d)  Maxwell electromagnetism, which clearly does not give a realistic description of a superconducting film. In this paper we shall consider a more realistic non-relativistic microscopic model where the electromagnetic field extends in three spatial dimensions. As pointed out in \cite{Hansson:2004fk}, this changes qualitatively the screening properties of the superconducting state compared to that implied by a two-dimensional electromagnetism. Furthermore, the surface plasmons are not gapped in this case which means that the nature of the phase, \ie whether it is topologically ordered, has to be critically reexamined. 

Technically, we proceed by deriving an effective low-energy action for thin, fluctuating superconducting films taking into account the realistic three-dimensional (3d) electromagnetic interaction among the charged carriers using the pseudo QED  (PQED) approach, which incorporates the effects of 3d electromagnetism in a non-local 2d action \cite{Marino:1993uq}. 
In a derivative expansion, the leading topological contribution to the effective theory is the BF term, just as in idealized models coupled to 2d electromagnetism. In this case, however, the gauge PQED action is of the same order.  As a consequence, the screening behaviour in a superconducting film is very different from a pure 2d superconductor. 
In particular the  Meissner effect, characteristic of superconductivity, is modified and the magnetic field around a vortex shows a power-law rather than an exponential decay inside a superconducting film as originally found by Pearl \cite{pearl64}. Similarly, electric charges are only screened by a power law, and the braiding phase obtained when moving a charge around a distant vortex equals the topological value -1 only up to power-law decaying corrections. 
As opposed to the QED$_3$ case, surface plasmons are gapless and disperse as $\omega\sim \sqrt k$ at low momenta, just as in a two-dimensional metallic film \cite{Stern1967}. This raises the question whether, in a 2d film coupled to 3d electromagnetism, there is a sharp distinction between the superconducting phase and the 2d metal. 
We address this question first by showing that the expectation value of the the vortex creation operator vanishes just as in the 2d abelian Higgs model which is know to be topologically ordered.  
We then calculate the ground states on a torus to determine to what extent the superconducting phase can be considered as topologically ordered in spite of being gapless. Our conclusion in this respect is similar to the one obtained by Bonderson and Nayak in the case of a quantum Hall liquid coupled to 3d electromagnetism \cite{BN}. 


The paper is organized as follows: 
In section \ref{sect:model} we introduce our model and derive the effective low-energy theory. 
In section \ref{sect:plasma} we extract the source-free equations of motion and analyze solutions describing gapless 2d plasmons. 
In section \ref{sect:ivc} we derive the effective interaction between electric and magnetic sources. We get exact results for the static charge-charge, and vortex-vortex potentials. We also determine the statistical charge-vortex interaction in the relativistic limit, where the speed of sound equals the speed of light. 
Section \ref{sect:vortenergy} contains the calculation of the energy of an isolated vortex, and in
section \ref{sect:toporder} we investigate the signatures of topological order in a superconducting film.  
In the last section, \ref{sect:phases}, we give a short summary of the results, present our conclusions concerning topological order in superconducting films, and make some remarks about possible future extensions of this work.  

We will use the notation for the metric
$\eta_{\mu\nu}=(+,-,-)$, $\partial^2=\eta^{\mu\nu}\partial_\mu \partial_\nu=\partial_t^2-\partial_i^2$, and momenta
$k_\mu=-i \partial_\mu$, $k=|\mathbf{k}|$, $\eta^{\mu\nu}k_\mu k_\nu=\omega^2-k^2$. 
We often set $\hbar=c=1$.


\section{ The effective low-energy theory} \label{sect:model}

In this section we derive the effective low-energy theory for a thin, fluctuating, superconducting flat film modeled by a time-dependent 2d Ginzburg-Landau action $S_\phi = \int dt d^2 r\, \mathcal{L_\phi}$ with
\be{nonrel}
\mathcal{L_\phi}=  \phi^\dagger (i \partial_t - 2e A_t )\phi - \frac {1} {2m} |( i\pmb\nabla- \frac {2e} c \mathbf A )\phi|^2 -V(n),
\ee
where $n = \phi^\dagger \phi$ is the two-dimensional density of Cooper pairs, and $V(n)$ a potential that fixes the mean value of the number density to $\av n = \bar n$.
The Cooper pair bosonic field has the electric charge  $2e$ and mass $m$. Since the aim here is to describe a thin film, the characteristic length scale is the Pearl length \cite{pearl64}, $\lambda_P$ which is related to the thickness of the film $d$, and the London length in the 3d material $\lambda_L$, as 
\be{pearl} 
\lambda_P = \frac {2\lambda_L^2} d = \frac {mc^2} {2e^2 \bar n_3 d} = \frac {mc^2} {2 e^2 \bar n},
\ee
where $\bar n_3$ is the average 3d Cooper pair density. 

The charged 2d Cooper pair field, $\phi(\mathbf r, t)$ is coupled to a 3d dynamical electromagnetic potential $A_\mu(\mathbf r, z, t)$. Rather than using a theory with mixed dimensionality, we shall use the PQED formalism\cite{Marino:1993uq}, where the effect of 3d electromagnetism on  2d matter is captured by the following non-local Lagrangian for the 2d vector potential $A_\mu(\mathbf r, t)$
\be{PQED}
{\mathcal L}_{PQED} =  -\frac1 4 F_{\mu \nu}  \frac{2}{\sqrt{\partial^2}} F^{\mu\nu},  
\ee
where $\partial^2=\partial_t^2/c^2 -\partial_i^2$.
The vector potential is normalized as in 3d so that $e$ is the dimensionless electric charge (recall that we set $\hbar =1$). 
If the 2d system is embedded in a medium, one must introduce relevant electric and magnetic susceptibilities in this expression. 

To proceed we parametrize the  Cooper pair field as,
\be{order}
\phi=\sqrt n\,  e^{i\Theta} = \sqrt n \,  e^{i \theta} \xi
\ee
where $\theta$ is the regular part of the phase, and $\xi$ is a singular phase factor describing point like vortices, such that the vortex current is given by 
$j_v^\mu =\frac i {2\pi} \epsilon^{\mu\nu\sigma} \partial_\nu (\xi^\star \partial_\sigma \xi)$.


The next step is to expand Eq. \eqref{nonrel} to quadratic order in the density fluctuation $\delta n = n - \bar n$ and integrate out $\delta n$. The quadratic terms in the resulting Lagrangian are
\be{semirel}
\mathcal{L_\phi}= \frac {c^2} {\lambda_P} \left[ \frac{1}{c_s^2}\left(\frac 1 {2e} \partial_t \theta - \frac 1 e a_t + A_t \right)^2 - \left(\frac 1 {2e} \pmb\nabla\theta - \frac 1 e \mathbf a + \frac 1 c \mathbf A \right)^2 \right] - A_\mu j_q^\mu \, ,
\ee
where we introduced a gauge field  $a_\mu =i \xi^\star \partial_\mu \xi/2$ dual to the vortex current,  the speed of sound  $c_s^2 = \bar n V^{\prime\prime} (\bar n)/m$, and 
also  added a coupling to an external charged current $j^\mu_q$ that describes quasiparticles. 

Following Ref. \cite{kvorning2014}, we linearize the quadratic terms in \eqref{semirel}  by introducing an auxiliary current three-vector $J^\mu$,
\be{lineraized}
\mathcal{L_\phi}= -\frac {\lambda_P} {4 c^2} \big( c_s^2 J^t J_t+J^i  J_i \big)  + J^t (\frac 1 {2e} \partial_t \theta - \frac 1 e a_t + A_t )+J^i  (\frac 1 {2e} \nabla_i \theta - \frac 1 e a_i + \frac 1 c  A_i ) - A_\mu j_q^\mu \, .
\ee
The equation of motion for $\theta$ implies the conservation law $\partial_\mu J^\mu = 0$, so we can parametrize the current as 
\be{}
J^\mu = \frac{e}{\pi } \epsilon^{\mu\nu\rho} \partial_\nu b_\rho \, .
\ee
Introducing the electric and magnetic fields, $e_i =\partial_t b_i - \partial_i b_t$ and $b= \epsilon^{ij} \partial_ib_j$, and using $j_v^\mu=\epsilon^{\mu\nu\rho}\partial_\nu a_\rho/ \pi$, 
the effective Lagrangian becomes,
\be{nrelfin}
\mathcal{L_\phi} = \frac{e^2 \lambda_P}{4\pi^2 c^2}  (\mathbf e\cdot \mathbf e - c_s^2 b^2) + \frac e {\pi} \big(A_t b-\frac 1 c \epsilon^{ij} A_i e_j \big) - b_\mu j_v^\mu - A_\mu j_q^\mu \, .
\ee 
In a later section we shall use a relativistic version of this model model, where $c_s = c=1$, and then the above expression simplifies to 
\be{relfin}
\mathcal{L}_{\phi, rel}=- \frac{e^2 \lambda_P}{8\pi^2 c^2} f_{\mu\nu} f^{\mu\nu} + \frac e {\pi} \epsilon^{\mu\nu\sigma} b_\mu\partial_\nu A_\sigma  - b_\mu j_v^\mu - A_\mu j_q^\mu \, ,
\ee
where $f_{\mu\nu} =\partial_\mu b_\nu - \partial_\nu b_\mu$. 
In the absence of a vortex source $j_v^\mu$, it is straightforward to integrate out the auxillary gauge field $b_\mu$ and get the effective electromagnetic response Lagrangian,
\be{massterm}
{\mathcal L}_A = \frac 1 {\lambda_P} \left[\frac{c^2}{c_s^2} A_t^2 - \mathbf A\cdot \mathbf A \right] - A_\mu j_q^\mu \, ,
\ee
which in the relativistic case simplifies to the usual mass term $\sim A_\mu A^\mu$. 

The model that we shall investigate in this paper is the sum of the PQED Lagrangian ${\mathcal L}_{PQED} $ in \pref{PQED} and  the matter Lagrangian given either by $\mathcal{L_\phi} $ in \pref{nrelfin} or ${\mathcal L}_A $ in \pref{massterm}.

%
%

\section{Plasma oscillations} \label{sect:plasma} 
Consider first a superconducting film without vortices. 
In the absence of quasiparticle sources (\ie  $j_q = 0$),  ${\mathcal L}_A $ reduces to a mass term,  
and varying ${\mathcal L}_A + {\mathcal L}_{PQED}$ with respect to $A_0$ and $A_i$ we get the following equations of motion in Fourier space
\be{eom1}
\frac 1 {\sqrt{k^2-\omega^2/c^2}} \left( k^2 A_t - \frac \omega c \mathbf k\cdot \mathbf  A \right)+ \frac {c^2} {\lambda_P c_s^2} A_t       = 0 \, , 
\ee
\be{eom2}
\frac 1 {\sqrt{k^2-\omega^2/c^2}} \left( \frac {\omega^2} {c^2} \mathbf A  - \frac \omega c \mathbf k A_t - k^2\mathbf A + (\mathbf k\cdot\mathbf A) \mathbf k \right)- \frac 1 {\lambda_P} \mathbf A = 0 \, .
\ee
To look for wave solutions, we first consider the possibility of a spatially transverse wave. In this case $\mathbf k \cdot \mathbf E =0$ which implies $c k^2 A_t = \omega \mathbf k\cdot\mathbf A$ which inserted in \eqref{eom1} gives $A_t = 0$. Using this  in \eqref{eom2} we get
\be{disptr}
\sqrt{k^2-\frac{\omega^2}{c^2}}+\frac{1}{\lambda_P}=0 \, .
\ee
This equation has no solutions.

We now turn to the spatially longitudinal mode. 
First, we assume $\omega \neq ck$ and multiply both \eqref{eom1} and \eqref{eom2} with $\sqrt{ -\omega^2/c^2 + k^2}$. Now taking the scalar product of \eqref{eom2} with $\mathbf k$, and combining the result with \eqref{eom1}, we get the following dispersion relation for the longitudinal mode,
\be{displong}
-\omega^2+c_s^2 k^2+\frac{c}{ \lambda_P} \sqrt{-\omega^2+c^2 k^2}=0
\ee
with the gapless real solution
\be{}
\omega=\left\{
                \begin{array}{ll}
                  c k & k\ll \frac{1}{\lambda_P}, \\
                  \sqrt{ \frac{2 c^2}{\lambda_P} k} &  \frac{1}{\lambda_P} \ll k \ll \frac{c^2}{\lambda_P c_s^2} \, , \\
                  c_s k& k\gg \frac{c^2}{\lambda_P c_s^2}.
                \end{array}
              \right.
\ee
This result illustrates a general finding that at low momenta the plasmon in a superconductor is gapless and essentially  indistinguishable from a plasmon in a metal \cite{Anderson1958, Anderson1963, Ohashi1998}.



\section{Interactions between vortices and charges} \label{sect:ivc}
In this section we consider the interaction between the excitations: quasielectrons and vortices.  We shall  in turn treat the three cases, quasielectron - quasielectron, vortex - vortex, and quasieletron - vortex. The first two have a non-trivial static limit, while the third is a velocity dependent charge - current interaction. 
Since our  theory  is quadratic in the gauge fields $A_\mu$ and $b_\mu$, we can integrate these fields out and compute the current-current interactions between the excitations. The algebra simplifies considerably in the relativist limit $c_s = c=1$ and we shall work in this regime in the following. Notice that for the static regime, where $\omega =0$, this yields the exact  result and we will assume that  it gives qualitatively correct results also for the charge - current interaction.

By varying the Lagrangian $\mathcal{L}_{rel}={\mathcal L}_{\phi, rel}+ {\mathcal L}_{PQED} $ with respect to $b_\mu$ and $A_\mu$, we get  in the Lorenz gauge ($\partial^\mu A_\mu=\partial^\mu b_\mu=0$) 
in Fourier space
\be{EOMF}
 \left(
\begin{array}{cc}
 2 \sqrt{k^2-\omega^2} \eta^{\mu\nu} & \frac i \pi \epsilon^{\mu \sigma \nu } k_\sigma, \\
 \frac i \pi \epsilon^{\mu \sigma \nu } k_\sigma & \gamma (k^2-\omega^2) \eta^{\mu\nu}
 \end{array}
\right) \left(
\begin{array}{c}
 A_\nu  \\
 b_\nu \\
\end{array}
\right)=
\left(
\begin{array}{c}
 j^\mu_q  \\
 j^\mu_v  \\
\end{array}
\right)\, ,
\ee
where we introduced $\gamma=\lambda_P/(2\pi^2)$ and for simplicity set $e=1$.
We now invert the matrix and get 
\be{eom}
\left(
\begin{array}{c}
 A_\mu  \\
 b_\mu \\
\end{array}
\right)=\frac{1}{\sqrt{k^2-\omega^2}+\sigma}
\left(
\begin{array}{cc}
 \frac{ \eta_{\mu\nu}}{2} & -\frac {i} {2 \pi}  \frac{\epsilon_{\mu\rho \nu} k^\rho}{\gamma (k^2-\omega^2)} \\
-\frac {i} {2 \pi}  \frac{\epsilon_{\mu\rho \nu} k^\rho}{\gamma (k^2-\omega^2)}  & \frac{ \eta_{\mu\nu}}{\gamma \sqrt{k^2-\omega^2}}
 \end{array}
\right)
\left(
\begin{array}{c}
 j^\nu_q  \\
 j^\nu_v  \\
\end{array}
\right),
\ee
where $\sigma^{-1}=\lambda_P$. 
By substituting  this solution into the Lagrangian $\mathcal{L}_{rel}$ we finally obtain  the action
\be{SREL}
S_{rel}=-\frac{1}{2} \int \frac{d\omega}{2\pi} \frac{d^2 k}{(2\pi)^2}
\left( \begin{array}{cc}
j^\mu_q  &  j^\mu_v  \\
\end{array}
\right) _{-\omega, -\mathbf{k}}
\frac{1}{\sqrt{k^2-\omega^2}+\sigma}
\left(
\begin{array}{cc}
 \frac{ \eta_{\mu\nu}}{2} & -\frac {i} {2 \pi}  \frac{\epsilon_{\mu\rho \nu} k^\rho}{\gamma (k^2-\omega^2)} \\
-\frac {i} {2 \pi}  \frac{\epsilon_{\mu\rho \nu} k^\rho}{\gamma (k^2-\omega^2)}  & \frac{ \eta_{\mu\nu}}{\gamma \sqrt{k^2-\omega^2}}
 \end{array}
\right)
\left(
\begin{array}{c}
 j^\nu_q   \\
 j^\nu_v  \\
\end{array}
\right)_{\omega, \mathbf{k}}.
\ee
\subsubsection{Charge - charge interaction}
By taking $\omega =0$ in the diagonal terms, we can read off the static potentials. For the charge - charge interaction we have
\be{}
V_q(k)=\frac 1 2  \frac{1}{k+\sigma}
\ee 
which in position space is
\be{}
V_q(r)=\frac 1 2  \int \frac{d^2 k}{ (2\pi)^2}\frac{\exp(i \mathbf{k}\cdot \mathbf{r})}{k+\sigma}=\frac 1 2 \int_{0}^{\infty} \frac{dk }{2\pi} \frac{k J_0(k r)}{k+\sigma}.
\ee
This integral can be performed analytically, see for example appendix of \cite{Clem2004}. At short distances $\sigma r \ll 1$, the potential diverges as $V_q\sim r^{-1}$, while at large distances $\sigma r \gg 1$ one finds $V_q \sim r^{-3}$. 
 Contrary to a bulk three-dimensional superconductor, where due  to the Higgs mechanism charges are screened exponentially, in a superconducting film screening is less effective and falls as a power law at large distances.

\subsubsection{Vortex -  vortex interaction}
Similarly, in the static limit the vortex - vortex interaction is given by
\be{}
V_v(k)= \frac{1}{\gamma k(k+\sigma)}
\ee
which in position space translates into
\be{}
V_v(r)=\int \frac{d^2 k}{ (2\pi)^2}\frac{\exp(i \mathbf{k}\cdot \mathbf{r})}{\gamma k (k+\sigma)}=\int_{0}^{\infty} \frac{dk }{2\pi} \frac{ J_0(k r)}{\gamma(k+\sigma)}.   
\ee
This integral can also be evaluated analytically\cite{Brandt2009}. At short distances the potential is logarithmic $V_v\sim \log (1/ r)$, while at large distances it decays as a power law $V_v\sim 1/r$. 
This result goes back to Pearl \cite{pearl64}, who was the first to study vortex-vortex interaction in thin superconducting films. We observe that the Pearl length $\lambda_P$ separates the logarithmic short-distance behavior of the potential from the power-law decay at large distances. For completeness, we also give the magnetic field far away from  a single vortex,  
\be{vortexfield}
B(r) \xrightarrow[r \to \infty ]{}   \frac {\phi_0} {2\pi} \frac {\lambda_P} {r^3},
\ee
where $r$ is the distance from the center of the vortex and $\phi_0= 2\pi {\hbar c}/{2e}$ is the superconducting  flux quantum, which in the units used in this section is $\phi_0 = \pi$. The result \pref{vortexfield} is in agreement with what is obtained in Ref. \onlinecite{fetter}.


\subsubsection{Vortex - charge interaction}
Finally, we  discuss the statistical vortex-charge interaction that gives the mutual statistics between vortices and charges.
The relevant quantity is the expectation value of two non-intersecting Wilson loops giving the space-time histories of a quasiparticle and a vortex,
\be{wilson}
\av{W_q[C_1] W_v[C_2]} = \exp [ i \oint_{C_1} dx_\mu  \oint_{C_2} dy_\nu G_{qv}^{\mu\nu}(x-y)  ],
\ee
where the pertinent Greens function is the Fourier transform of the Euclidean version of the off-diagonal element in \eqref{SREL},
\be{offgfcn}
G_{qv}^{\mu\nu}(x) = -\frac 1 {2\pi} \int \frac {d^3 p} {(2\pi)^3}    \frac {e^{ipx}} {p + \sigma} \frac {\epsilon^{\mu\nu\rho} p_\rho} {\gamma p^2}.
\ee
Here we introduced the Euclidean 3-momentum $p^\mu$ and $p=\sqrt{p_0^2+p_i^2}$.
A direct calculation gives,
\be{offgfcn2}
G_{qv}^{\mu\nu}(x) =   \frac {\epsilon^{\mu\nu\rho} x_\rho} {4 |x|^3} \left[ 1 + \frac 1 \pi C(\sigma r) \right ],
\ee
where $C(x) =  2 x^2 J^\prime (x)$ with
$$
J(x) = \frac 1 x \int_0^\infty dy\, \frac {\sin (y)} {y + x} \,
$$
which can be evaluated analytically.
The first term in the parenthesis in \eqref{offgfcn2} gives, 
\be{linking}
\av{W_q[C_1] W_v[C_2]} = e^{i \pi L[C_1,C_2] }
\ee
where the linking number $L[C_1,C_2]$ is a topological invariant which counts the number of times the particle encircles the vortex, and thus
correctly gives their mutual $\pi$-statistics. The second term gives a  correction to this phase, but since   $\lim_{x\rightarrow \infty} C(x) = -4/x$ this correction is negliable for large loops that, although linked, never come close together. This shows that the braiding statistics of quasiparticles and vortices is only defined up to power-law corrections in thin films that are described by the PQED-Higgs model. In contrast, if we were to take a 2d Maxwell term, instead of the PQED Lagarangian, we would get an exponentially suppressed correction to the $\pi$ phase \cite{hansson89}. 


\section{Vortex energy} \label{sect:vortenergy}
In the subsequent discussion of the nature of the superconducting phase, the energy gap to topologically non-trivial excitation, \ie vortices, will be of 
importance. In this section we calculate the energy of a single vortex from the 2d relativistic model $\mathcal{L}_{rel}={\mathcal L}_{\phi, rel}+ {\mathcal L}_{PQED} $. 

Since the PQED Lagrangian is nonlocal in time, we shall not attempt to derive a Hamiltonian using canonical methods, but rather obtain the energy-momentum tensor by varying the action with respect to the metric tensor. A direct calculation gives
\be{emtensor}
T_{\mu\nu}^{\mathrm PQED} = - F_{\mu\alpha} \frac 2 {\sqrt{\partial^2}} F_\nu^\alpha + \frac 1 4 \eta_{\mu\nu} 
F_{\alpha\beta}  \frac 2 {\sqrt{\partial^2}} F^{\alpha\beta} - \frac 1 2 F_{\alpha\beta}  \frac {\partial_\mu\partial_\nu} {\partial^{3/2}} F^{\alpha\beta}.
\ee
For a static configuration we get (in an obvious notation), 
\be{T00}
T_{00}^{\mathrm PQED} = \mathbf E_A \cdot \frac 1 {\sqrt{-\nabla^2}} \mathbf E_A + B_A\cdot \frac 1 {\sqrt{-\nabla^2}}  B_A.
\ee
The topological $bdA$ does not depend on the metric and thus gives no contribution to $T_{\mu\nu}$. The Maxwell term for the $b_\mu$ potential gives (in units $c=e=1$)
\be{T00a}
T_{00}^{\mathrm \phi} = \frac { \lambda_P} {4\pi^2 } \mathbf{E}_b \cdot \mathbf{E}_b.
\ee
The total energy density is $T_{00}=T_{00}^{\mathrm PQED}+T_{00}^{\mathrm \phi}$.

Taking now a static point vortex, Eq. \pref{eom} immediately gives the momentum space expressions,
\be{bfield}
B_A(k) &=&  \frac 1 {2\pi\gamma} \frac 1 {k+\sigma},  \\
\mathbf E_b (\mathbf{k}) &=& \frac i \gamma \frac {\mathbf k} {k(k+\sigma)}.
\ee
The vortex energy  is now obtained by substituting this into the above expressions for $T_{00}$ and integrating over the two-dimensional film in Fourier space
\be{energy}
\begin{split}
E_v &= \int \frac{d^2 k} {(2\pi)^2} \left( B_A(k) \frac 1 k B_A(k) + \frac {\lambda_P} {4\pi^2} \mathbf{E}_b(\mathbf{k})\cdot \mathbf{E}_b(\mathbf{-k}) \right) \\
&= \frac \pi 2 \frac 1 {\lambda_P} \int_0^{\lambda_P/\xi} \frac {dp} {1+p} =  \frac \pi 2 \frac 1 {\lambda_P} \ln \left(1 + \frac {\lambda_P} \xi \right),
\end{split}
\ee
where we introduced a short-distance cutoff $\xi$ that has a natural interpretation as a correlation length. Restoring the charge $e$, this expression for the energy is precisely the one given in Ref. \onlinecite{degennes} (notice that our $\lambda_P$ differ by a factor $8\pi$ from the parameter $\lambda_{eff}$ in this reference). From this result, and that in section \ref{sect:plasma}, we see that not only static correlation functions, but also energetics and collective dynamics is captured correctly by our effective 2d theory.


\section{Signatures of topological order }   \label{sect:toporder}
As discussed in some detail in \oncitec{Hansson:2004fk} a \emph{fluctuating} superconductor is topologically ordered and cannot be characterized by  a non-zero expectation value of a local gauge-invariant order parameter; note that the Cooper pair field $\phi$ is not gauge-invariant and by Elitzur theorem averages to zero in a fluctuating superconductor.\footnote{ This does of course not mean that a superconductor is  defined only by its topological properties. On the contrary, the most significant characteristics are related to transport and screening, and of particular importance is the Meissner effect, that clearly distinguishes a superconductor from a metal. In 3d superconductors, the Meissner effect means that applied magnetic fields penetrate a superconductor only over a distance of the London length, $\lambda_L$. In the 2d toy model with a 2d Maxwell term, the situation is the same. All these characteristics are however exhibited also by a non-fluctuating superconductor where the kinetic term $\sim E^2$ for the electromagnetic field is ignored, implying a description as a charged superfluid with a spontaneously broken $U(1)$ symmetry. }
To decide whether the thin superconducting film discussed in this paper is topologically ordered, we must therefore use signatures that  directly 
probe the phase structure without assuming the existence of a local order parameter. With this in mind, we shall in this section first use the formalism introduced by   't Hooft to classify the phases of gauge theories and then discuss the ground state degeneracy on higher genus surfaces. 

\subsection{The vortex operator and its correlators} \label{sect:algandvort}
In Ref. \cite{thooft} 't Hooft showed that the phases of a gauge theory in three space-time dimensions are characterized by the ground state expectation values of a pair of operators, $\mathcal{A}_C$ and $\mu(\mathbf x)$. The operator $\mathcal{A}_C$ is the Wilson loop defined on the closed curve $C$, while $\mu$ is a local operator that implements aa gauge transformation that is is singular at the point $x$. These operators satisfy the equal time commutation relations
\be{thooftalg}
\mu (\textbf{x}, t) {\mathcal A}_C(t)   =  e^{i\pi w[C,\textbf{x}] } {\mathcal A}_C(t)     \mu (\textbf{x}, t) \, ,
\label{thooft2}
\ee
where $w[C,\textbf{x}]$   the winding number which counts how many times the curve $C$ winds around the point $\textbf{x}$.
A two-dimensional gauge theory admits several possible phases: In the confining phase, $\av \mu \neq 0$ and $\av{ \mathcal{A}_C} = \exp (-\sigma A(C))$, where $A(C)$ is the area of the loop $C$. In the Higgs, \ie  superconducting, phase $\av \mu \ne 0$ while $\av {\mathcal{A}_C }= \exp (-\gamma L(C))$, with $L$ is the length of the loop. The third possibility is a  gapless Coulomb phase with both $\mu \ne 0$ and a perimeter law for the Wilson loop. As shown by Polyakov, (2+1)D compact QED is confining \cite{polyakov} and, as already discussed, the (2+1) abelian Higgs model is in a Higgs phase. 

To determine which phase describes the PQED-Higgs model we shall calculate $\av \mu$ using the vortex quantization techniques developed in Refs. \cite{marino1990solitons,marino1988quantum, marino1995new} (early studies on vortex quantization can also be found in Refs. \cite{xx,xxx1,xxx2,xxx3,xxx4,xxx5}, and for a detailed review, see \cite{marino2017qftcmp}). We first show how to define a vortex creation operator, and then calculate its two-point function from which we can extract $\av \mu$.

\subsubsection{The vortex creation operator}  \label{sect:vortices}

We define an operator that, when acting on the vacuum, creates eigenstates of the  magnetic flux operator
\begin{eqnarray}
\hat\Phi= \int d^2x \hat j_v^0 = \int d^2x \epsilon ^{ij}\partial_i \hat A_j \, ,
\label{30}
\end{eqnarray}
where $\hat A$ is the electromagnetic quantum field; in the rest of this section we shall suppress the hats for ease of notation. After introducing the quantum vortex creation operator $\mu(\textbf{x},t)$, we have
\begin{eqnarray}
\Phi |\mu\rangle = \Phi \Big[ \mu(\textbf{x},t) \ket 0 \Big ]=  \phi_0 |\mu\rangle\, ,
\label{31}
\end{eqnarray}
An operator with the above property is \cite{marino1988quantum}
\begin{eqnarray}
\mu(\textbf{x},t)  =  \exp \left \{i \phi_0
\int_{\textbf{x},L}^\infty d\textbf{z}^i \epsilon^{ij} \Pi^j(\textbf{z}, t) \right \}\, ,
\label{31a}
\end{eqnarray}
where $\phi_0$ is the superconductor flux quantum,  $L$ is a contour that starts at the point $\mathbf{x}$ and goes to infinity, and $\Pi^i =\partial \mathcal{L}/\partial \dot{A}_i$ is the momentum canonically conjugate to $A^i$, 
satisfying the canonical commutation relation
\begin{eqnarray}
[A^i(\textbf{x},t) , \Pi^j(\textbf{y}, t) ] =i \delta^{ij} \delta^2(\textbf{x}-\textbf{y})\, .
\label{31b}
\end{eqnarray}
One can show \cite{marino1988quantum} that 
\begin{eqnarray}
 \mu (\textbf{x},t) A^i(\textbf{y},t) = \left[ A^i(\textbf{y},t) + \frac{\phi_0}{2\pi}
  \partial^i_{(y)} \arg(\textbf{y}-\textbf{x}) \right ] \mu (\textbf{x},t)\, . 
\label{31c}
\end{eqnarray}
It can be also shown that for the total phase $\Theta$ and the Cooper pair field $\phi$ defined in section \ref{sect:model}, one gets 
\begin{eqnarray}
 \mu (\textbf{x},t) \Theta(\textbf{y},t) = \left[\Theta (\textbf{y},t) +  \frac{\phi_0}{2\pi}   \arg(\textbf{y}-\textbf{x}) 
 \right ] \mu (\textbf{x},t) \, ,
\label{31d}
\end{eqnarray}
and 
\begin{eqnarray}
 \mu (\textbf{x},t) \phi(\textbf{y},t) =  \exp \left\{i\frac{\phi_0}{2\pi}   \arg(\textbf{y}-\textbf{x})\right \}  \phi(\textbf{y},t) 
  \mu (\textbf{x},t) .
\label{31e}
\end{eqnarray}

Applying the operation $\oint_C d\textbf{y} \cdot \mathbf{\nabla}_{\textbf{y} }$ on both sides of \eqref{31d} and noting that the regular part $\theta$ of the total 
phase $\Theta$ does not contribute to the integral, we get
\be{thooft1}
[ \mu (\textbf{x}, t), \oint_C d\textbf{y} \cdot  \textbf{a} (\textbf{y} ,t)] = \frac{\phi_0}{2\pi}  \oint_C d\textbf{y} \cdot 
\mathbf{\nabla} \arg(\textbf{y}-\textbf{x}) \mu (\textbf{x}, t) \, \equiv \phi_0\  w[C,\textbf{x}]\   \mu (\textbf{x}, t)
\ee
where we recalled that $ \textbf{a}(\textbf{x},t) =  \pmb{ \nabla} \Theta (\textbf{x},t)$ and
\begin{equation}
w[C,\textbf{x}]=\frac{1}{2\pi} \oint_C d\textbf{y} \cdot \mathbf{\nabla} \arg(\textbf{y}-\textbf{x}) 
\end{equation}
is  the winding number which counts how many times the curve $C$ winds around the point $\textbf{x}$.
We now introduce the 't Hooft operator, 
\begin{equation}
{\mathcal A}_C(t) = \exp\Big\{ie\oint_C d\textbf{y} \cdot \textbf{a}(\textbf{x},t)\Big\} \, ,
\end{equation}
and using the commutator  \eqref{thooft1} along with the Baker-Hausdorff formula for ${\mathcal A}^{-1}_C \mu {\mathcal A}_C$, we finally get 
the t' Hooft commutation relation \eqref{thooftalg}.



\subsubsection{The vortex two-point function}
In this subsection we calculate the two-point Euclidean vortex correlation function from which we can extract $\av\mu$.
General vortex correlation functions can be obtained directly from the expression (\ref{31a}), or, alternatively, by treating the vortex operator as a disorder variable, in the sense of Kadanoff and Ceva \cite{kadanoff1971determination,marino1988quantum,marino1995new}.

For a general theory with an action depending on a $U(1)$ field strength tensor $F_{\mu\nu}$, the following vortex  two-point correlation function was derived  in Refs. \onlinecite{marino1988quantum} and \onlinecite{marino1995new} 
\begin{eqnarray}
\langle \mu(x) \mu^\dagger(y)\rangle &=& Z_0^{-1} \int DA_\mu  
\exp \left \{- S\left[F^{\mu\nu}+ \tilde B^{\mu\nu}  \right] \right \}\, ,
\label{32}
\end{eqnarray}
where we used the shorthand notation $x\equiv (\textbf{x}, x^0)$ \etc for the position in Euclidean spacetime, and where the external field $ \tilde B^{\mu\nu}$ is given by 
\begin{eqnarray}
 \tilde B^{\mu\nu}(z; x, y) = \phi_0 \int_{x,L}^y \epsilon^{\mu\nu\alpha} \delta^3(z-\xi) d\xi_\alpha \, .
\label{33}
\end{eqnarray}
In this expression, $L$ is an arbitrary curve connecting the points $x$ and $y$ in Euclidean space-time. It can be shown that the above correlation function is $L$-independent \cite{marino2017qftcmp}, despite the explicit dependence of the external field (\ref{33}) on $L$.


In Lorenz gauge, the effective action obtained from the relativistic theory derived in section \ref{sect:model} is,
\begin{eqnarray}
 S= \frac{1}{4} \int d^3x \, F^{\mu\nu}\Big [\frac{ M + 2(-\Box)^{1/2} }{(-\Box)}  \Big ] F_{\mu\nu}\, ,
\label{6axx}
\end{eqnarray}
where $\Box$ is the Laplace operator in Euclidean spacetime and $M\equiv 2/\lambda_P$ has dimension of inverse length. 
Substituting  \eqref{6axx}  into  \eqref{32} to get,
\begin{eqnarray}
\langle \mu(x) \mu^\dagger(y)\rangle &=& Z_0^{-1} \int DA_\mu  
\exp \left \{- \frac{1}{4} \int d^3x \Big [F^{\mu\nu}+ \tilde B^{\mu\nu}  \Big ]
\left[\frac{ M + 2(-\Box)^{1/2} }{(-\Box)}\right]\Big [F^{\mu\nu}+ \tilde B^{\mu\nu}  \Big ] \right \} \, ,
\label{6xx}
\end{eqnarray}
and then performing the functional integral over $A_\mu$, we obtain
\begin{eqnarray}
\langle \mu(x) \mu^\dagger(y)\rangle_{} &=&
\exp \left \{\Lambda(x,y;L) 
-\frac{1}{4}\int d^3z
 \tilde B^{\mu\nu}  \left [\frac{ M + 2(-\Box)^{1/2} }{(-\Box)}\right  ] 
\tilde B_{\mu\nu}  \right \} \, .
\nonumber \\
\label{569a}
\end{eqnarray}
where $\Lambda(x,y;L) $ is 
\begin{eqnarray}
\Lambda(x,y;L) &=&\frac{1}{8} 
\int d^3z d^3z'
\tilde B^{\mu\nu}(z) \tilde B^{\alpha\beta} (z') P^{\mu\nu}_\lambda P^{\alpha\beta}_\rho
 \left [\frac{ M + 2(-\Box)^{1/2} }{(-\Box)}\right  ]
 D^{\lambda\rho}(z-z') 
\left [\frac{ M + 2(-\Box)^{1/2} }{(-\Box)} \right  ],
\nonumber \\
 \label{570a}
\end{eqnarray}
with $P^{\mu\nu}_\lambda=\partial^\mu\delta^\nu_\lambda-\partial^\nu\delta^\mu_\lambda$ and
$D^{\lambda\rho}(z-z')$  the gauge field propagator, given by
\begin{eqnarray}
D^{\lambda\rho} =\frac{\delta^{\lambda\rho}}{\left[ M + 2(-\Box)^{1/2}\right]} + {\rm gauge\  terms} \, .
\label{571a}
\end{eqnarray}
Inserting (\ref{571a}) in (\ref{570a}), 
\begin{eqnarray}
 \left [\frac{ M + 2(-\Box)^{1/2}}{(-\Box)} \right  ]&\times&\left [\frac{1}{ M + 2(-\Box)^{1/2}} \right  ]
\times  \left [\frac{ M + 2(-\Box)^{1/2}}{(-\Box)} \right  ]  
=\left [ \frac{2}{(-\Box)^{3/2}} + \frac{M}{(-\Box)^2}  \right  ]
\label{572a}
\end{eqnarray}
and using  the inverse Fourier transforms
\begin{eqnarray}
&& \mathcal{F}^{-1}\left[  \frac{1}{(k^2)^{3/2}}\right ]=\lim_{m\rightarrow 0} \frac{K_0(m|x|)}{2\pi^2} = -\frac{1}{2\pi^2} \left( \ln \frac{m |x|}{2}+\gamma_E\right),
\nonumber \\
&& \mathcal{F}^{-1}\left[ \frac{1}{k^4} \right ] =\lim_{m\rightarrow 0} \frac{e^{-m |x|}}{8\pi m}= \frac{1}{8\pi }
\left(\frac{1}{m}-|x| \right),
\label{576a}
\end{eqnarray}
where $m$ is an $U(1)$ breaking infrared mass regulator,   \eqref{572a} gives the position space expression
\begin{eqnarray}
\left [ \frac{2}{(-\Box)^{3/2}} + \frac{M}{(-\Box)^2}  \right  ]\to F(x-y)=
\lim_{m\rightarrow 0}\left [-\frac{1}{\pi^2} \left( \ln \frac{m |x-y|}{2}+\gamma_E\right) 
+ \frac{M}{8\pi }
\left(\frac{1}{m}-|x-y| \right)\right ].
\nonumber \\
\label{572ab}
\end{eqnarray}
Inserting (\ref{572ab}) and (\ref{33}) in  (\ref{570a}) and substituting the result in (\ref{569a}),  one finds that the second term in (\ref{569a}) is cancelled, and we are left with the $L$-independent term (for a detailed derivation in a closely related problem, see \cite{Marino1995,marino2017qftcmp})
\begin{eqnarray}
\langle \mu(x) \mu^\dagger(y)\rangle = \exp \left[ \phi_0^2 \left\{F(x-y) - F(\epsilon) \right \} \right]  = \frac{ \epsilon^\nu  }{|x-y|^\nu}
 \exp \left \{- \mathcal{M} |x-y|\right\} \, ,
\label{578a}
\end{eqnarray}
where $\epsilon$ is a short-distance cutoff, 
%
and where $ \mathcal{M} =\frac{M\phi_0^2}{8\pi}$, and $\nu=\frac{\phi_0^2}{\pi^2}$. Note that this correlation function does not depend on the IR cutoff $m$. 
The short-distance cutoff $\epsilon$ can be 
absorbed by introducing the  renormalized vortex creation operator 
$$
\mu_R(x) \equiv \mu(x) \epsilon^{-\nu/2} .
$$
In terms of this operator, we finally get
\begin{eqnarray}
\langle \mu_R(x) \mu_R^\dagger(y)\rangle = \frac  1 {|x-y|^\nu} \exp{ \left \{- \mathcal{M} |x-y| \right \} } \, .
\label{578aa}
\end{eqnarray}
This should be compared with the corresponding result for the 2d abelian  Higgs model with the Maxwell term.  In the Higgs phase one finds \cite{Marino1995,marino2017qftcmp}
\begin{eqnarray}
\langle \mu_R(x) \mu_R^\dagger(y)\rangle_{AHM} = \exp{ \left \{-\mathcal{\tilde M} |x-y|+\frac{\gamma}{|x-y|} \right \} } \, ,
\label{578x}
\end{eqnarray}
which again decays exponentially at large distances, and  differs essentially from \eqref{578aa} only at short distances, and in both cases we have $\av\mu=0$. 
Note the difference with the results in Section \ref{sect:ivc}, where we saw that the large distance screening of both electric and magnetic fields differed  qualitatively in the Higgs phases of QED and PQED, with only the former showing  exponential screening.

We end this section with three remarks: \\ 

i) From \eqref{578aa} it is natural to interpret ${\mathcal M} = {\phi_0^2}/{(4\pi \lambda_P)}$ as the vortex mass, and it is pertinent to ask how this is compatible with the result \eqref{energy} which was also derived from a relativistic model. We notice that the logarithmic dependence on correlation length is absent. One should however remember that the operator $\mu$ was constructed as to create a local topological charge eigenstate, which from the outset does not depend on any length scale. Although the original Ginzburg-Landau model has stable mean field solutions describing vortices, this is not necessarily true for the pure gauge model ${\mathcal L}_A + {\mathcal L}_{PQED}$ where the only scale is $\lambda_P$. To find such solutions one must introduce a vortex source as in \ref{sect:vortenergy}. It is an open question if and how one could retain the information about the correlation length in the pure gauge theory description of the matter sector.\\​

ii) It is not hard to show that the dependence on the infrared regulator $m$ cancels in   any correlation function with zero total vorticity, 
while it remains in those with non-zero vorticity. The latter vanish as the $U(1)$ invariance is retained in the $m\rightarrow 0$ limit. For example
$$
\langle \mu_R(x) \mu_R(y)\rangle =  \exp \left[ \phi_0^2 \left\{-F(x-y) - F(\epsilon) \right \} \right]  \propto \lim_{m\rightarrow 0} m^\nu e^{-\frac{\mathcal{M}}{m}} \rightarrow 0 \, . \\
$$
iii) In the unbroken phase, where the superconducting condensate density vanishes, the vortex correlation function for the abelian Higgs model was given in Ref. \cite{Marino1995,marino2017qftcmp},
\begin{eqnarray}
\langle \mu_R(x) \mu_R^\dagger(y)\rangle_{AHM} = \exp{ \left \{\frac{\gamma}{|x-y|} \right \} }
\underset{|x-y|\rightarrow \infty} \longrightarrow  1 \, ,
\label{578xy}
\end{eqnarray} 
This  implies $\langle 0|\mu\rangle =1$, which means the vortex operator actually does not create any genuine excitations, because these should   be orthogonal to the vacuum. This is consistent with the description of the vacuum of the unbroken phase as a vortex condensate with $\av \mu \ne 0$. 
It is interesting that the behavior of the correlator is qualitatively different in PQED.
Taking $M \rightarrow 0$ in \eqref{578aa} yields 
\begin{eqnarray}
\langle \mu_R(x) \mu_R^\dagger(y)\rangle =
\frac 1 {  |x-y|^\nu }
\underset{|x-y|\rightarrow \infty} \longrightarrow 0  \, .
\label{578aax}
\end{eqnarray}
This corresponds to  a ``soft'' phase  where $M=0$ but also $\langle \mu(x)\rangle =0$, see Ref. \cite{RKEM1983}. The vortex two-point correlation function would in this case have a power-law decay, in accordance with \eqref{578aax}, so the quantum vortex excitations are gapless.
Interestingly, this is not a conventional gapless phase, since for arbitrary values of $\nu$, there is a cut in the propagator, rather than a pole; a behaviour which is reminiscent of that  in the 2D Tomonaga-Luttinger model \cite{TL,LP}.  This is of course not a description of a metallic film since it assumes the existence of an, admittedly strongly fluctuating, pairing field. It could possibly describe a thin film in a pseudogap phase of preformed Cooper pairs.



\subsection{Ground state degeneracy on a torus }   \label{sect:gsdeg}

\begin{figure}[ht]
\begin{center}
\includegraphics[width=0.5\textwidth]{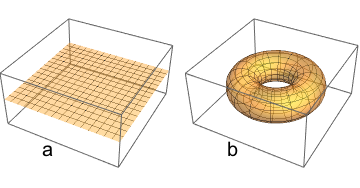}
\caption{(a) Flat and (b) curved embeddings of $T^2$ into $T^3$.}\label{fig1}
\end{center}
\end{figure}

One of the hallmarks of topological order is the ground state degeneracy on surfaces with non-zero genus. 
For a superconductor with two-dimensional electromagnetism that is described by a pure BF theory, there are four ground states on a torus $T^2$, corresponding to the four possible ways of inserting $Z_2$ fluxes for the statistical gauge fields $A$ and $b$ \cite{Hansson:2004fk}. Here we investigate the  ground state degeneracy problem of a two-dimensional superconductor coupled to electromagnetism that lives in three dimensions.  Before getting into technical details we notice that there are different ways how one can set-up the problem and embed a two-dimensional torus $T^2$ into a three-dimensional torus $T^3$. In Fig. \ref{fig1} we plot a flat and curved $T^2$ embedded into $T^3$. In the following we will only consider the flat embedding (Fig. \ref{fig1} a) because in this case adding $Z_2$ magnetic fluxes costs no bulk energy. In contrast, in the case of the curved torus (Fig. \ref{fig1} b) an insertion of a $Z_2$ flux comes with a finite 3d bulk energy cost, which necessarily lifts the ground states degeneracy. Here we analyze the problem using two approaches: In the first we closely follow the work in Ref. \cite{BN} on quantum Hall liquids coupled to three-dimensional electromagnetism, and in the second we attempt to use the PQED formalism developed in the earlier sections. 

\subsubsection{Bonderson-Nayak method} 
Consider a spacial 3-torus with size $L$ in the $x y$-plane defined by the superconducting film, and $L_z$ in the third direction. We take the Lagrangian as the sum of the 2d relativistic Lagrangian $\mathcal{L}_{\phi, rel}$ in  \eqref{relfin} and the usual three-dimensional Maxwell term,
\be{mixedact}
L = - \frac{e^2 \lambda_P}{8\pi^2 } \int d^2r\,  f_{\mu\nu} f^{\mu\nu} + \frac e {\pi}  \int d^2 r\,  \epsilon^{\mu\nu\sigma} b_\mu\partial_\nu A_\sigma  
- \frac 1 4 \int d^3r\, F_{MN} F^{MN},
\ee
where $\mu, \nu=t,  x, y$ and $M, N=t,  x, y, z$. $F_{MN}$ is the field strength corresponding to the vector potential $A_M$. In general, the presence of the superconducting film breaks translational invariance in the $z$-direction, but in the ground state both the charge density and the charge current vanish which effectively restores the translation invariance along $z$ axis. Now we Fourier decompose the gauge potentials as
\be{gfnorm}
\begin{split}
b_\mu (\mathbf r, t)  &=\frac 1 L \sum_{\mathbf k} e^{i \mathbf{k} \cdot \mathbf{x}} b_\mu(\mathbf{k},t),   \\
A_M (\mathbf r, z, t) &= \frac 1 {L L_z^{1/2} } \sum_{\mathbf k,k_z}  e^{i(\mathbf{k} \cdot \mathbf{x}+k_z z)}A_M(\mathbf{k}, k_z, t), 
\end{split}
\ee
where the normalization was chosen as in \cite{BN}. 
In the gauge $A_0 = b_0 = 0$ the Lagrangian for the spatial zero modes ($\mathbf{k}=0$, $k_z=0$) becomes
\be{zmlag}
L = \frac {e^2 \lambda_P} {4\pi^2} \dot b_i^2  + \frac e \pi \frac 1 {L_z^{1/2}} \epsilon^{ij} \dot A_i b_j + \frac 1 2 \dot A_i^2+\frac 1 2 \dot A_z^2,
\ee
where $i,j=x, y$.
Now we integrate out $\dot A_i$ and get (after dropping the $\dot A_z^2$ term)
\be{zmlag2}
L_{eff} = \frac m 2  \dot b_i^2 - \frac K 2 b_i^2,
\ee
where $m=e^2\lambda_P/(2\pi^2)$ and $K= e^2/(\pi^2 L_z) $. This is a 2d harmonic oscillator with frequency $\omega^2 = 2/(\lambda_P L_z) = (1/\lambda_L)^2 (d/L_z$). Restoring $\hbar$ and $c$, we get the energy gap for zero-mode solutions
\be{zgap}
\Delta E= \hbar \omega = \frac{\sqrt{2}\hbar c} {\sqrt{ \lambda_P L_z}}. 
\ee
We thus see that for a fixed Perl length $\lambda_P$ the gap \eqref{zgap} scales as $1/L_z^{1/2}$. We note that taking a scaling limit, $L_z \rightarrow \infty$, $\lambda_P\rightarrow 0$ with $\lambda_P L_z\rightarrow \text{const}$, $\Delta E$, remains finite.

What about finite momentum modes? In a finite $T^2$ that is embedded in a finite $T^3$ they are gapped as well. In particular, the plasmon modes studied in Sec. \ref{sect:plasma} have a gap $\Delta E_{pl}\sim 1/\sqrt{L}$ if $L\ll \lambda_P$ and $\Delta E_{pl}\sim 1/L$ if $L\gg \lambda_P$. In addition, we expect that photons that propagate along $z$-direction have a gap $\Delta E_{ph,z}\sim 1/L_z$.\footnote{
To the extent that these photons completely decouple from the superconductor, the state defined by the scaling limit  $\lambda_P L_z\rightarrow \text{const}$, would be topologcially ordered in the conventional sense. This limit might not be completely unrealistic if the electromagnetic field is screened at some finite distance and we consider strongly type II materials where $\lambda_L$ is very small. }

In summary, for finite $L_z$ all excitations have an energy gap that scales as an inverse power-law of the system size while for sufficiently large $L\gg \lambda_P$ the four ground states are expected to split only exponentially $\sim \exp(-L/\lambda_P)$. As a result, the superconducting film studied in this paper is a quasi-topologically ordered phase of matter using the terminology introduced by Bonderson and Nayak \cite{BN}.

\subsubsection{The PQED method}
Here we attempt to apply the PQED formalism.
We convert the zero-mode Lagrangian \eqref{zmlag} (dropping $\dot A_z^2$ term) to its PQED counterpart by the replacement
\be{repl}
\frac 1 2 \dot A_i ^2 \rightarrow    \dot A_i   \frac 1 {\sqrt{\partial_t^2}}  \dot A_i.
\ee
Using now the two-dimensional normalization for the Fourier transform
\be{gfnorm}
\begin{split}
b_\mu (\mathbf r, t)  &=\frac 1 L \sum_{\mathbf k} e^{i \mathbf{k} \cdot \mathbf{x}} b_\mu(\mathbf{k},t),   \\
A_\mu (\mathbf r, t) &= \frac 1 {L  } \sum_{\mathbf k}  e^{i(\mathbf{k} \cdot \mathbf{x})}A_\mu(\mathbf{k}, t)
\end{split}
\ee
and integrating out $\dot A_i$ gives
\be{cmlag3}
L^{PQED}_{eff} = \frac {e^2 \lambda_P} {4\pi^2} \dot b_i^2  - \frac 1 2  \frac {e^2} { \pi^2} b_i \sqrt{\partial_t^2}\,  b_i.
\ee
The last term is non-local and it is not obvious how to solve this quadratic problem.
If we ignore possible problems related to  $\partial_t^2$ being a negative definite operator and take the naive square root $\sqrt{\partial_t^2} \to \partial_t$,
the second term in Lagrangian becomes a total derivative and can be ignored. Thus we are left with a mass less theory consistent with taking the large $L_z$ limit of \eqref{zmlag2} at finite $\lambda_P$. We expect that for a finite $L_z$ the non-local PQED gauge Lagrangian must be modified, but that is beyond the scope of the present work.



\section{Summary, conclusions and outlook}   \label{sect:phases}

In this paper we proposed a non-local version of the Ginzburg-Landau model as an adequate low-energy description a thin  2d superconducting film. 
We derived the corresponding effective gauge theory which consists of a non-local PQED action for the electromagnetic field $A_\mu$, a Maxwell term for 
the gauge field $b_\mu$ describing the supercurrent and a BF term which couples the two gauge fields. 
Using this effective theory, we derived the interaction potentials between charges and vortices,  the surface plasmon dispersion relation, and the energy of a vortex, and verified that they agree with earlier results. We also found that the braiding phase for a  charge encircling a vortex gives the expected minus sign, up to power-law corrections. 

Thus convinced that the gauge theory gives a proper description of the thin superconducting film, we use it to determine to what extent this system can be considered as topologically ordered. The immediate answer would be no, since the plasmon gap vanish. Note however, that keeping a finite transverse size (which might well be the correct thing to do in a realistic system), the plasmon gap remains and so does the ground state degeneracy. A superconducting film thus exhibits quasi-topological order in the sense of Bondeson and Nayk \cite{BN}. 
We also studied the quantum vortex correlation function and showed that its long-distance behaviour is the same as for the 2d abelian Higgs model, and in particular found that $\av\mu=0$ for the vortex creation operator, which satisfies the 't Hooft algebra with the Wilson loop. From this we would conclude that a thin superconducting film is in a similar phase as the 2d abelian Higgs model, which is known to be topologically ordered. 

There are several directions in which this work could be extended: 
 At a technical level it is a challenge to carry out the calculations of the braiding phase for the non-relativistic case. One can also wonder whether the calculation of the vortex correlation function can be modified to reproduce correctly the logarithmic dependence of the vortex energy on the correlation length. Conceptually it would be interesting to attempt a dual formulation of the low-energy effective theory; for a self-dual theory of a somewhat related mixed-dimensional problem, see \cite{Hsiao2017}. Another potentially interesting direction is to apply the ideas developed here to layered superconductors.


\vskip 3mm\noindent {\bf Acknowledgements}
THH and ECM are grateful to Cristiane de Morais Smith for the hospitality at the University of Utrecht where this work was initiated.
This work was supported in part by CNPq (Brazil), Faperj (Brazil), by the Brazilian Government Program ``Science without Borders'', and by the Swedish Research Council. 
SM is supported by the Emmy
Noether Programme of German Research Foundation
(DFG) under grant No. MO 3013/1-1.

%


\begin{thebibliography}{20}

\bibitem{wen1990topological} X.-G. Wen, Int. J. Mod. Phys. B4, 239 (1990).

\bibitem{bala92} A. P. Balachandran and P. Teotonio-Sobrinho, Int. J. Mod. Phys. A 8, 723 (1993).
       
\bibitem{blau1991topological} M. Blau and G. Thompson, Ann. of Phys. 205, 130 (1991).

\bibitem{Moroz2017} S. Moroz, A. Prem, V. Gurarie and L. Radzihovsky, \prb, 95, 014508 (2017).

\bibitem{Hansson:2004fk} T. Hansson, V. Oganesyan, and S. Sondhi, Ann. of Phys. 313, 497 (2004).

\bibitem{Marino:1993uq} E. C. Marino, Nucl. Phys. B408, 551 (1993).

\bibitem{pearl64} J. Pearl, App.  Phys. Lett., 5, 65 (1964).

\bibitem{Stern1967} F. Stern,  Phys. Rev. Lett. 18, 546  (1967).

\bibitem[Bonderson \& Nayak(2013)]{BN} Bonderson, P., \& Nayak, C.\ 2013, \prb, 87, 195451.

\bibitem{kvorning2014} T. Kvorning, {\it Adding Majorinos to Superconductors}, Licentiate thesis, Stockholm University, 2014.

\bibitem{Anderson1958} P. W. Anderson, Phys. Rev. 112, 1900 (1958).

\bibitem{Anderson1963} P. W. Anderson, Phys. Rev. 130, 439 (1963).

\bibitem{Ohashi1998} Y. Ohashi, S. Takada, J. Phys. Soc. Jap. 67, 551 (1998).

\bibitem{Clem2004} J. R. Clem, J. Supcond. 17, 613 (2004).

\bibitem{Brandt2009} E. H. Brandt, Phys. Rev. B 79, 134526  (2009).

\bibitem{fetter} A. L. Fetter and P. C. Hohenberg, Phys. Rev. 159, 330 (1967).

\bibitem{hansson89} J. Grundberg, T. H. Hansson, A. Karlhede and U. Lindstr\"om, Phys. Lett. B. 218, 321  (1989).

\bibitem{degennes} P. G. De Gennes, {\it Superconductivity of metals and alloys}, Addison-Wesley (1989).

\bibitem{thooft} G. 't Hooft, Nucl. Phys. B 138, 1 (1978).

\bibitem{polyakov} A. M. Polyakov, Nucl. Phys. B 120, 429 (1977).

\bibitem{marino1990solitons} E. C. Marino, in {\it Applications of Statistical and Field Theory Methods to Condensed Matter}, A. Bishop et al eds., Springer, pp. 121-140 (1990).

\bibitem{marino1988quantum} E. C. Marino, Phys. Rev. D38, 3194 (1988).

\bibitem{marino1995new} E. C. Marino, Int. J. Mod. Phys. A10, 4311 (1995).

\bibitem{xx} J. Fr\"ohlich and P. A. Marchetti, Lett. Math. Phys. 16, 347 (1988); Commun. Math. Phys. 121, 177 (1989).

\bibitem{xxx1} G. W. Semenoff and P. Sodano, Nucl. Phys. B328, 753 (1989).

\bibitem{xxx2} M. L\"uscher, Nucl. Phys. B326, 557 (1989).  

\bibitem{xxx3}  R. Jackiw and S. Y. Pi, Phys. Rev. D42, 3500 (1990). 

\bibitem{xxx4}  A. Kovner, B. Rosenstein and D. Eliezer, Nucl. Phys. B350, 325 (1991) ; Mod. Phys. Lett. A5,  2733 (1990).

\bibitem{xxx5}  V. F. M\"uller, Z. Phys. C51, 665 (1991). 

\bibitem{marino2017qftcmp} E. C. Marino, {\it Quantum Field Theory Approach to Condensed Matter Physics}, Cambridge University Press (2017).

\bibitem{kadanoff1971determination} L. P. Kadanoff and H. Ceva, Phys. Rev. B3, 3918 (1971).


\bibitem{Marino1995} E.~C. Marino,\ Int. J. of Mod. Phys. A, 4311 (1995). 

\bibitem{RKEM1983} R. K\"oberle and E. C. Marino, Phys. Lett. B126, 475 (1983).

\bibitem{TL} S. Tomonaga, Progr. in Th. Phys. 5, 544 (1950); J. M. Luttinger, J. of Math. Phys. 4, 1154 (1963).

\bibitem{LP} A. Luther and I. Peschel, Phys. Rev. B12, 3908 (1975).

\bibitem[Hsiao \& Son(2017)]{Hsiao2017} W.-H.  Hsiao and D. T. Son, \prb, 96, 075127 (2017). 

%


\end{thebibliography}
\end{document}